\documentstyle[aps,prc,prl,epsf,multicol]{revtex}

\newcommand{\AmS}{{\protect\the\textfont2
  A\kern-.1667em\lower.5ex\hbox{M}\kern-.125emS}}
\begin{document}
\draft
\title { Sub MeV Particles Detection and Identification in the MUNU detector }
\author{
The MUNU collaboration\\
M. Avenier$^1$, 
C. Broggini$^3$,
J. Busto$^2$,
C. Cerna$^3$,
V. Chazal$^2$,
P. Jeanneret$^2$,
G. Jonkmans$^2$,\\
D.H. Koang$^1$\footnote{Corresponding author. Tel.: +33 4 7628 4048; fax: +33 4 7628 4004
\\e-mail koang@isn.in2p3.fr},
J. Lamblin$^1$, 
D. Lebrun$^1$,
O. Link$^4$,
R. L\"uescher$^2$,
F. Ouldsada$^4$,
G. Puglierin$^3$,\\
A. Stutz$^1$,
A. Tadsen$^3$,
J.-L. Vuilleumier$^2$} 
\address{$^1 $Institut des Sciences Nucl\'eaires, IN2P3/CNRS-UJF, 53 Avenue des Martyrs, F-38026 Grenoble, France\\  
$^2 $Institut de Physique, A.-L.  Breguet 1, CH-2000 Neuch\^atel, Switzerland\\
$^3 $INFN, Via Marzolo 8, I-35131 Padova, Italy\\ 
$^4 $Physik-Institut, Winterthurerstrasse 190, CH-8057 Z\"{u}rich, Switzerland}
\maketitle
\begin{abstract}
We report on the performance of a 1 m$^{3}$ TPC filled with CF$_{4}$ at 3 bar, immersed in liquid scintillator and viewed by photomultipliers. 
 Particle  detection, event identification and  localization achieved by measuring both the electric current and the scintillation light are presented. Particular features of $\alpha$ particle detection are also discussed. Finally, 
 the $^{54}$Mn photopeak, reconstructed from the Compton scattering and recoil angle is shown.\\
\end{abstract}
\begin{multicols}{2} 
\section*{\bf I. INTRODUCTION} 
	Large time projection chambers (TPC) have been used in high energy particle experiments as central trackers and more recently for identification of heavy nuclei and radioactive beams \cite{Bae98}, \cite{Hli98}.  
   At lower energy, gas TPC were operated both for particle tracking and calorimetry. A  TPC has been used in an experiment searching for $\mu$-e conversion \cite{Bry85}.
 In the MeV energy range, the TPC's involved are of smaller size. First double beta decay spectra were obtained by Elliot et al. \cite{Moe91} using a TPC of about 100 $l$. In the Gothard experiment \cite{Vui93}, a TPC  with an active volume of 200 $l$ provided one of the best limits on neutrinoless double beta decay in $^{136}$Xe. TPC with optical readout were also considered for dark matter searches \cite{BuK94},\cite{Tit98}. 
 	The MUNU collaboration has built a 1 m$^{3}$ gas TPC for the experimental study of $\overline{\nu}_{e}e^{-}$ scattering near a nuclear reactor in Bugey (France). The detector is designed to identify recoil electrons and reconstruct particle tracks well below 1 MeV. The tracking capability of the TPC (angle, threshold, and track containment) is essential for background rejection and allows online background measurements.  The TPC vessel is made of transparent acrylic and is installed inside a 10 m$^{3}$ light collection vessel, filled with liquid scintillator and viewed by phototubes. Both the electrec current and the scintillation light of particles interaction in the TPC gas can be collected. In the following section, a description of the detector is given whereas in section III the energy and efficiency calibrations of the detector are described. Waveforms and event identification are presented in section IV. In section V  the energy and angular resolutions are discussed. Incident gamma energy reconstruction is also presented.
 Finally, conclusions are given in section VI. 
\section*{\bf II. THE MUNU DETECTOR} 
\begin{figure}[ht]
\vspace{2.cm} 
\epsfysize=14cm 
\centerline{\epsffile{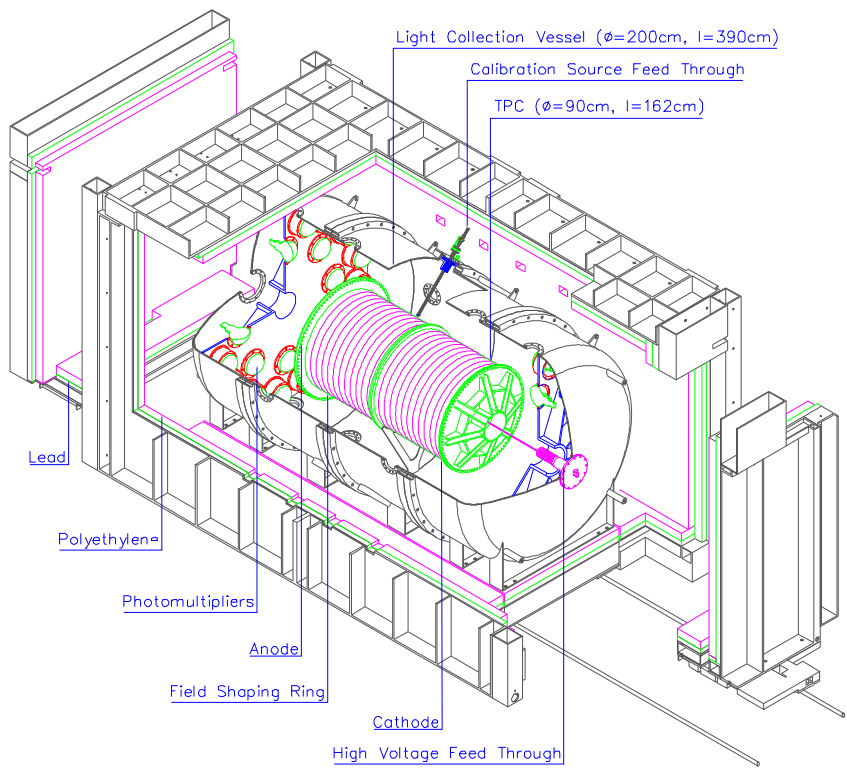}}
        \caption[]{Isometric sketch view of the MUNU
	detector inside its shieldings.}
       \label{fi:detector}
\end{figure}

  The general concept of the MUNU experiment and its detector have been presented previously \cite{Ams97}. Here we give an update and discuss the performances of the detector. The detector consists of a central TPC filled with CF$_4$ gas, inside a peripheral light collection vessel filled with 10 m$^{3}$ of liquid scintillator. It is equiped with photomultipliers and can act as veto for Compton and cosmic events. The detector shown in Figure \ref{fi:detector} is installed inside  low activity passive shieldings composed of borated polyethylene (8 cm thick) and lead (15 cm thick).   
\subsection*{II.1 The TPC} 
The 1 m$^3$ time projection chamber can be  filled with CF$_4$ gas at a pressure between 1 and 5 bar.  This gas was chosen because of its high density (1.06$\cdot$ 10$^{21}$ electrons per cm$^3$ at STP) and its low Z. This gives a high target density
combined with relatively little multiple scattering. Most data, in particular all those reported in this paper, have
been taken at a pressure of 3 bar, at which electron tracks can be reasonably well resolved down to 300 keV.

\begin{figure}[hbt]
\epsfysize=10cm
\centerline{\epsffile{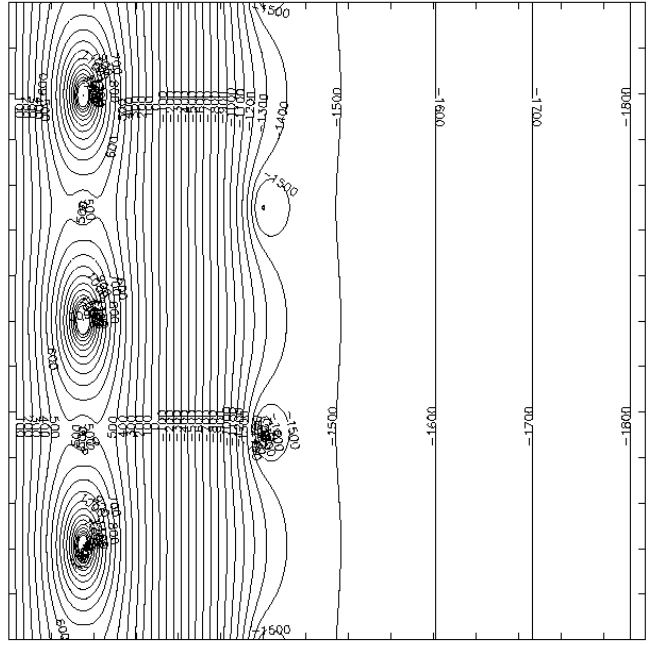}}
\caption{The electric field configuration in the vicinity of the read-out 
planes. The horizontal axis corresponds to the $z$ drift axis, the vertical 
axis is the bisecting line between the $x$ and $y$ pick-up strips. The 
ticks on both axis are every 2mm. The equipotentials are shown with the corresponding
voltages. The grid is at -2000 V, the anode wires at 3540 V, the potential
wires at 390 V.} 
\label{fi:TPC_Efield}   
\end{figure}

The gas is contained in a cylindrical acrylic vessel
of 90 cm inner diameter, and with an inner length of 162 cm.  A cathode made from  high purity electrolytic copper foil  is mounted on one lid. A read-out plane consisting, from inside to outside, of a grid, an anode plane and an x-y pick-up plane, is mounted on the other side. The arrangement is
shown in Figure \ref{fi:TPC_Efield}. The grid wires have a diameter of 100 $\mu$m. The distance between grid wires is 4.95 mm. The anode plane has 20 $\mu$m anode wires, separated by 100 $\mu$m potential wires. The distance between anode and potential wires  is also 4.95 mm.  The distance from the anode to the x-y plane is 3 mm whereas the grid and the anode plane are separated by a 8.5  mm gap. The spacing between x strips on one side of the supporting PET foil, or  y strips on the other side, is 3.5 mm, giving in total 256 x and 256 y strips. 

The drift field is defined by the cathode, the read-out plane, primarily the grid, and the field shaping rings mounted on the outside of the acrylic vessel. The x-y plane is grounded through the read-out preamplifiers. The drift time and drift velocity are optimized in order to get the best spatial resolution.  At 3 bar, we are operating the TPC with a drift field of 206 V/cm. This gives a drift velocity of 2.14 cm.$\mu$s$^{-1}$, and a total drift time of 74.8 $\mu$s. This was determined from the total drift time of muon tracks spanning the entire chamber, from cathode to grid. The voltages applied to the last field shaping ring, the grid, the anode and the potential wires must lead to sufficient amplification around the anode wires. Moreover they must provide a symmetric field configuration around the anode wires to minimize mechanical strain and sagging, and a homogeneous drift field all the way down to the grid over the entire area.  The Garfield code was used to find a configuration meeting these requirements. It is shown in Figure \ref{fi:TPC_Efield}.

Gas purity is essential for good charge collection, but also to minimize contributions to the background from radon emanations. The CF$_4$ gas is therefore constantly circulated at a flow rate of 500 l/h through a commercial high temperature getter made by SAES \cite{Saes} followed by a cold trap. The cold trap consists of a coil tubing followed by a a cell filled with 50 g of active charcoal. It is immersed in an ethanol bath kept at a temperature of 190 $^0$K. 

\begin{figure}[ht]
\epsfysize=8cm
\epsfxsize=12cm
\centerline{\epsffile{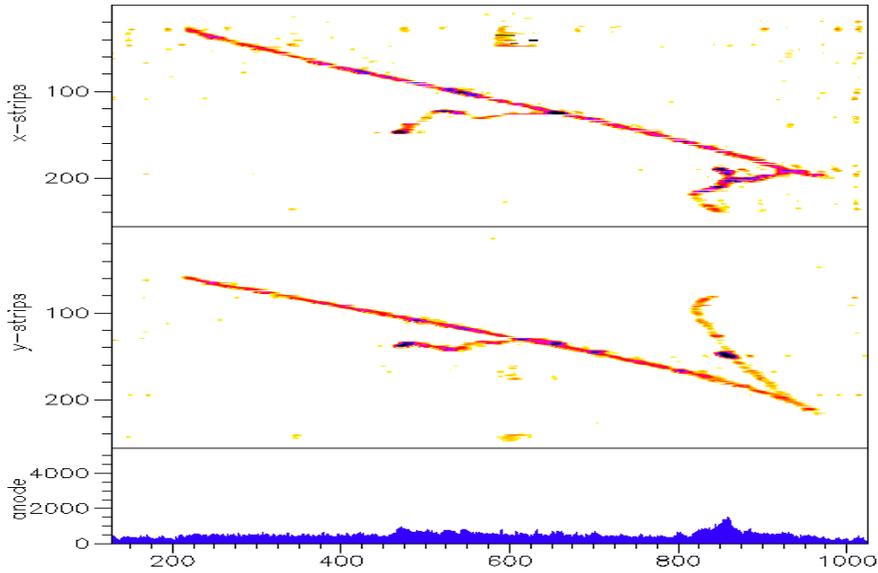}}
\caption
{A cosmic muon producing two high energy delta electrons. The one near the center with an energy of 1 MeV is contained, and the increased charge density at its end is clearly visible.}  
\label{fi:2460_11} 
\end{figure}

The signals from the x and y strips and the anode  
are amplified in current voltage preamplifiers and 
sampled at 12.5 MHz (80 ns sampling time) in 12 bit FADC's with logarithmic response. Electronics noise on the x and y strips is 
serious, because of the length, over 2 m,  of the flat signal cables from the TPC to the preamplifiers located outside the polyethylene, lead and steel shielding. A simple procedure allows to significantly reduce it. First a Fourier transform of the signals is  performed. Frequency peaks identified as noise sources are eliminated. The loss in spatial resolution is acceptable. To give an idea of the quality of the tracks, an example is shown in Figure \ref{fi:2460_11}.
\subsection*{II.2 The external light collection detector}
The TPC vessel is immersed in a cylindrical steel tank painted with TiO2 for diffuse reflection and filled with a mineral oil based liquid scintillator (NE235, provided by Nuclear Enterprise, Scotland). The scintillator has an attenuation length of 8 m at 430 nm and is viewed by 48 hemispherical photomultipliers of 20 cm diameter (EMI 9351 low activity series, provided by Hamamatzu Photonics), subdivided in two groups of 24 each on the anode and cathode sides. The mean quantum efficiency of the photocathodes is around 25 \%. The PMT's are sampled with the same FADC's as the TPC anode and x-y strips. Radioactive sources can be introduce into the light collection vessel at an axial distance of 26cm from the center of the detector, on the anode side through an acrylic pipe. The radial position can be as close as 1 cm from the TPC lateral wall. 

The main role of the external detector is to identify and to reject the Compton electrons induced by $\gamma$ rays coming from outside as well as from inside the TPC. Since the TPC walls are made of transparent acrylic, the scintillation light of  the CF$_{4}$ can also be collected. Within the TPC, the primary scintillation light of CF$_{4}$  
can be emitted anywhere while avalanche scintillation light is produced only in the vicinity of the anode plane during the amplification process.
The scintillation light spectrum of CF$_{4}$ has a narrow maximum at 160 nm and a broader one around 300 nm \cite{Van78}-\cite{Bro95}. The light is detected by the phototubes through the acrylic walls and the liquid scintillator. Due to the transmittance of the acrylic and of the liquid scintillator, only a fraction of about 10 \% of the spectrum with wavelengths larger than 400 nm  makes significant contribution to the light signal in our detector. 
 \subsection*{II.3 The readout system }
As already mentionned, anode, x-y strips and phototube signals are read out with  8 bits flash ADC's with logarithmic response. The memory size is 1024 words. The sampling interval is 80 ns and is given by a common clock. This sampling time is well ajusted to the electric pulses from the anode which have a relatively slow rise time of few hundred ns. Signal pulses from the phototubes associated with Compton electrons are much faster ($<$ 30 ns) and are shaped to meet the sampling frequency. The digitized data of the flash ADC's provide the pulse shape  of the signals delivered by the phototubes, the anode, and the x-y strips. This way, projection on the x-y planes can be reconstructed. Time bins can be converted to relative position coordinate along the TPC Z-axis using the drift velocity in the TPC (2.14 cm.$\mu$s$^{-1}$ in normal running conditions).
	
The scintillation light produced in the liquid scintillator or the primary scintillation light of the CF$_{4}$ which is strong enough to be seen for $\alpha$ particles and not for nearly minimum ionizing electrons and muons, are collected within one or two sampling bins of 80 ns. Charges arrive to the anode with a time delay proportional to their distances from the anode plane.  Absolute z position can thus be defined for events with prompt scintillation light.  For events such as confined single electrons, which have no primary scintillation light in CF$_{4}$ and no interaction with the liquid scintillator, only relative z positions can be obtained.\  
\section*{\bf III. Calibrations of the detector}  \  
 In the following, we first discuss the light collection and energy calibration for the liquid scintillator detector. In particular, we discuss the $\gamma$ ray  detection efficiency and we compare the measurements to simulations. We then describe the energy calibration  of the TPC.
\subsection*{III.1 Energy calibration and light collection in the external detector }
  In the light collection vessel, $\gamma$ rays undergo  multiple Compton scattering and are in most cases totally absorbed within the scintillator liquid. Radioactive sources ($^{54}$Mn, $^{135}$Cs, $^{22}$Na) are used for light collection and energy calibrations.
Along the cylindrical vessel axis (z), a convenient position parameter is the assymetry (i.e. difference-to-sum ratio) of light signals collected by phototubes of the cathode and anode sides. The light collection has been measured as function of this parameter using the total absorption peak of a $^{54}$Mn source.   

\begin{figure}[ht]
\vspace{1.cm}
	\epsfysize=8cm	
\centerline{\epsffile{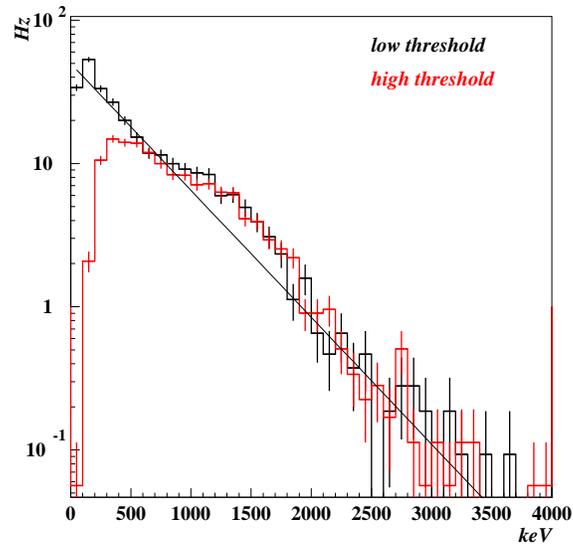}}
\caption{Total absorption spectra obtained with different radioactive sources measured in the external detector, and compared to simulations (solid lines).}	
        \label{fi:MnAC}
\end{figure}

 Experimental data  agree with simulations based on known emission spectrum and attenuation length of the liquid scintillator, quantum efficiency of the phototubes, and optical properties of other components. At the central part of the vessel, the response is quite homogeneous over a large region ($\pm$ 40 cm from the center along the detector axis) where about  7.5$\%$ of photons emitted in the liquid scintillator are collected.  At  80 cm  from the center, close to both ends of the TPC, the fraction of light collected is 9.5$\%$. It increases to 15 $\%$ at 120 cm from the center of the detector. Position dependence of the light collection can be corrected for an optimum energy resolution or exploited for event localization.
 The quantity of photoelectrons collected was also evaluated from total absorption spectra of $^{54}$Mn measured for each individual phototube and compared to single photoelectron spectra. For an energy deposition of 1 MeV  at the central part of the detector, 144 $\pm$ 8 photoelectrons were collected in the 48 phototubes while 150 were predicted.
Total absorption spectra obtained with different radioactive sources and corrected for position dependence are compared to simulations in Figure \ref{fi:MnAC}. The energy resolution is  13$\%$ at 1 MeV. In the following text, only corrected values of energy will be used.
\subsection*{III.2  $\gamma$'s detection efficiency}  
\begin{figure}[ ]
	\epsfysize=8cm	
\centerline{\epsffile{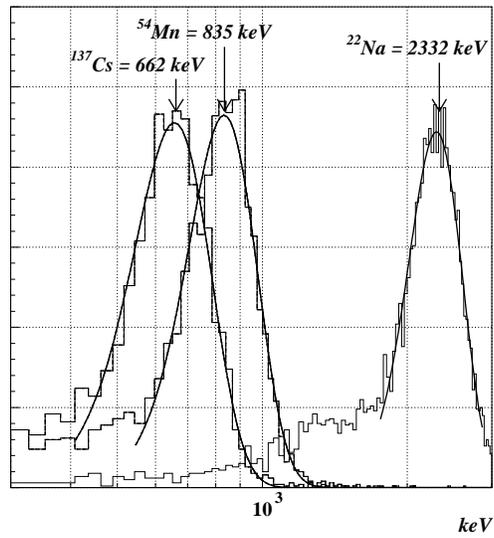}}	
\caption[]{Energy spectra collected in the 10 m$^3$ external detector at two hardward energy thresholds.}
        \label{fi:AC1}
\end{figure}

\begin{figure}[ht]
        \vspace{1.cm}
	\epsfxsize=8cm
	\epsfysize=8cm	
\centerline{\epsffile{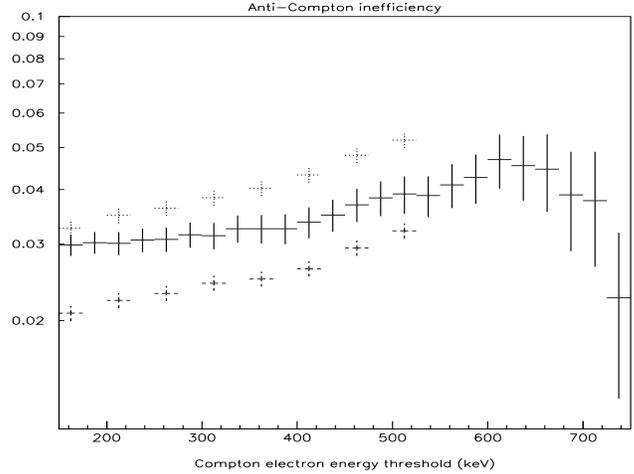}}
        \caption[]{Fraction of Compton electron events not detected by the external detector as a function of the recoil electron energy. Data measured with a $^{54}$Mn source (E$_\gamma$= 835 keV) are shown with error bars and are compared to simulations at two anti-Compton energy thresholds: 100 keV(dashed cross) and 125 keV(dotted cross).}
        \label{fi:eff}
\end{figure}

	CF$_{4}$ is composed of low Z atoms, only a small amount of very low energy $\gamma$ rays will lose the entire energy in the gas. Most of them undergo a single Compton scattering in the gas and are absorbed in the scintillator. The MUNU detector has been designed for an optimum detection of $\gamma$ rays associated with a Compton electron inside the TPC. To minimize $\gamma$ absorption, the lateral walls of TPC are made of very thin (5 mm) acrylic. At an energy, threshold set at 100 keV which corresponds to a total collection of 15 photoelectrons, the contributions of dark noise from individual phototubes is not negligeable. For this reason a multiplicity criteria asking for at least 5 phototubes hit is required for a trigger. The counting rate for the 10 m$^3$ of liquid scintillator within the shielding is 710 s$^{-1}$ including 270 s$^{-1}$ coming from cosmic muons. The energy spectrum measured at $\sim$ 100 keV (low) threshold shown in Figure \ref{fi:AC1} corresponds to a counting rate of 0.05 s$^{-1}$ kg$^{-1}$ of liquid scintillator.
   Signals from the scintillator can be applied to collect or to veto TPC events associated with a scattered $\gamma$ ray. (Similarly, a signal at an energy threshold of $\sim$ 8 MeV electron equivalent can be applied to veto cosmic muons).
    The detection efficiency of scattered $\gamma$ has been measured by using a $^{54}$Mn source of 15 kBq inserted at 1 cm from the TPC wall. Electron energy spectra were collected with and without the anti-Compton veto operating at 100 keV hardware threshold. The anti-Compton efficiency is evaluated from the ratio of the two electron spectra, corrected from background contributions measured in the absence of the radioactive source.
     Only 3$\%$ of the $\gamma$ associated with a Compton electron of more than 300 keV escape detection. The scattered photon is lost when it does not interact with the liquid scintillator or when it is absorbed by inert material inside the vessel(acrylic walls, cables). Figure \ref{fi:eff} shows the comparison of the experimental results to simulations at two thresholds: 100 and 125 keV \cite{Lam00}.
 For a $\gamma$ ray coming from outside the light collection vessel, the inefficiency is much smaller, since it has to travel twice through about 50 cm of liquid scintillator. Taking also into account the solid angle and the absorption due to the liquid scintillator, the simulation shows that the probability 
to miss a $\gamma$ ray leaving more than 300 keV in the TPC is less than  10$^{-6}$. 
\subsection*{III.3 Energy calibration of the TPC}  
\begin{figure}[ht]
	\vspace{1.cm}
	\epsfxsize=8cm
	\epsfysize=8cm	
\centerline{\epsffile{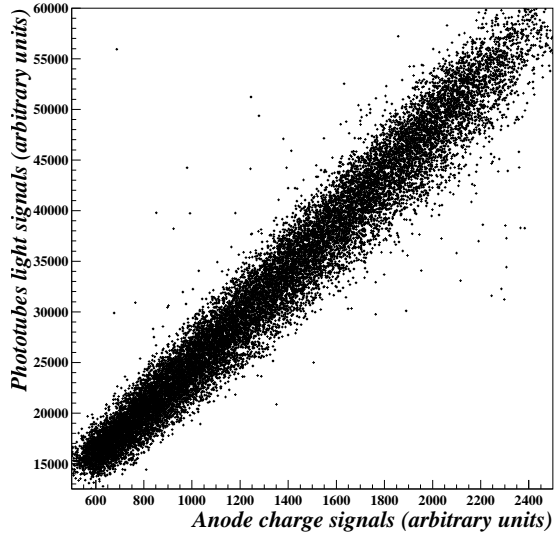}}	
        \caption[]{Event per event comparison of the amplitudes of the current and light signals.}
        \label{fi:TPCAC}        
\end{figure}

\begin{figure}[ht]
\vspace{1.cm}
	\epsfxsize=8cm
	\epsfysize=8cm	
\centerline{\epsffile{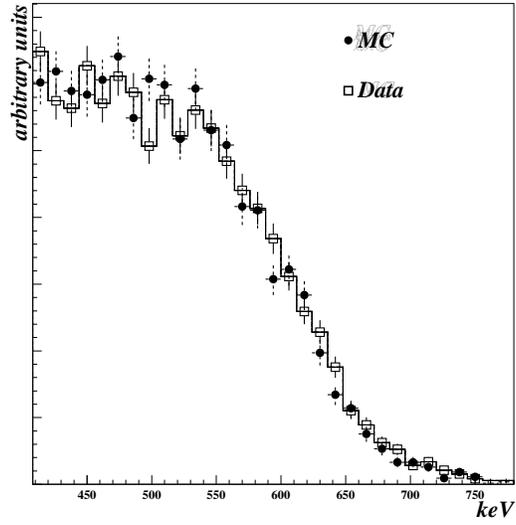}}
\caption[]{A Compton spectrum of $^{54}$Mn in the TPC, compared to Monte Carlo simulation.}
        \label{fi:MnTPC}	
\end{figure}

\begin{figure}[ ]
\vspace{1.cm}
	\epsfxsize=8cm
	\epsfysize=8cm	
\centerline{\epsffile{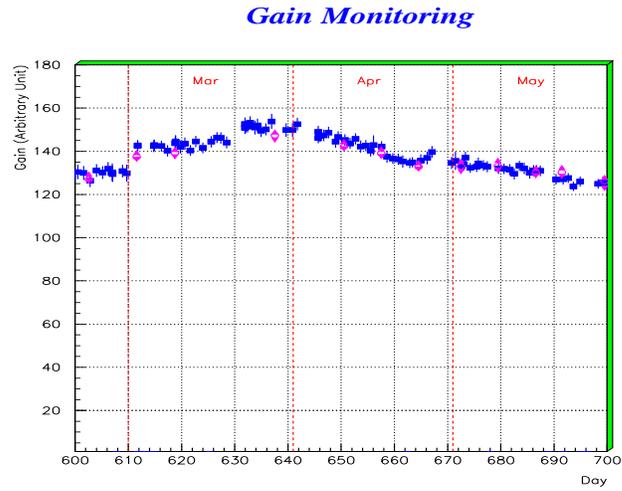}}
\caption[]{Gain monitoring with radioactive sources(diamonds) and muons( squares).}
        \label{fi:gainmoni}	
\end{figure}

\begin{figure}[ht]
\vspace{1.cm}
	\epsfxsize=8cm
	\epsfysize=8cm	
\centerline{\epsffile{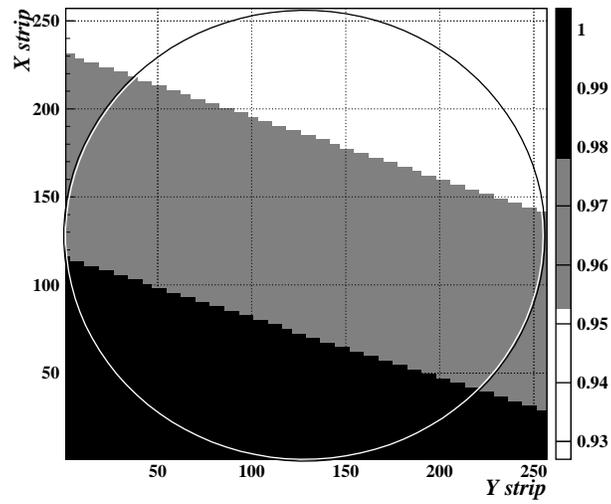}}
\caption[]{Gain varaiation on the anode plane. 
The vertical scale on the right indicates percentage of gain variation.}
        \label{fi:gainxy}	
\end{figure}

	 Particle tracks in the TPC produce charges which drift to the anode plane. There the charge is amplified and avalanche light is produced.  Light emission during the multiplication phase in CF$_{4}$ has been previously observed\cite{Pan95}. In our detector, about 2.10$^4$ photoelectrons are collected per MeV of electron.
	 Electric current signals are more sensitive to electromagnetic noises and have larger pedestal fluctuations. Current and light waveforms from the same electron event will be shown in the following section. Avalanche light signals have better bin to bin pedestal stability and are used instead of the electric current signals for energy evaluation and for determination of track ends.  An event per event comparison shows the linear correlation between the amplitudes the two types of signals as given in Figure \ref{fi:TPCAC}.    
	Compton electron spectra were obtained using a $^{54}$Mn source both from the electric current and the light and are compared to simulations in Figure \ref{fi:MnTPC}.   Mean energy loss of muon events per unit length in the TPC is used to monitor day-to-day gain variations (Figure \ref{fi:gainmoni}).  The gain changes on the anode plane are mapped with alpha and muon events and are presented in Figure \ref{fi:gainxy}.

\section*{\bf IV. Waveforms and event identification}
\subsection*{IV.1 Muon events} 
\begin{figure}[ht]
	\epsfxsize=8cm
	\epsfysize=10cm	
	\centerline{\epsffile{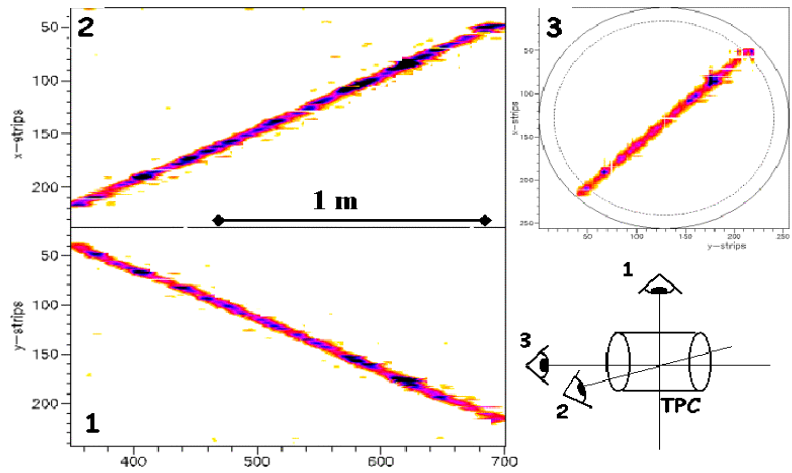}}
      \caption[]{A muon event: the xz, yz and xy projections are shown.}
      \label{fi:muon}
\end{figure}

	The experimental area in Bugey has an overhead shielding in heavy concrete equivalent to 20 m of water. Compared to ground level, the cosmic muon flux is reduced by a factor of five. The muon counting rate is expected to be 65 s$^{-1}$ in the TPC and  270 s$^{-1}$ in the light collection vessel. 
 	A cosmic muon attaining the TPC has to cross more than 100 cm of liquid scintillator. With a mean energy lost of about 1.8 MeV/cm, it leaves a very large prompt signal in the phototubes. This signal can be used as trigger to select muon events or to veto muon correlated events.
	The energy loss of cosmic muons in the CF$_{4}$ covers an energy range from hundred keV to a few MeV.  The muon mean energy loss per unit length in the TPC can be used to continuously monitor the gain variations in a complementary way to the measurements with radioactive sources.
 Muon events crossing both cathode and anode have  x-y projection confined within the TPC radius and are used to evaluate the drift velocity. The time difference between the beginning and the end of the track correponding to a maximum drift path of 1.62 m.
The x-z, y-z projections of a muon event are shown together with 
the x-y projection in Figure \ref{fi:muon}. 
\subsection*{IV.2 Alpha particles }  
\subsubsection*{IV.2.1 Primary light emission }  			
  Alpha particles from natural radioactivity have a range shorter than 1 cm in CF$_{4}$ at 3 bar : 4.1 mm at 5 MeV and 8.6 mm at 8 MeV. The spatial extensions of $\alpha$ events are limited to a few x-y strips and a few time bins.
    While no primary scintillation light was seen for Compton electrons at a threshold as low as 30 keV electron equivalent,  the light from $\alpha$ particles was detected. A prompt signal equivalent an 153 keV electron (22 photoelectrons) is produced in the liquid scintillator by a  5.3 MeV  $\alpha$ particles. In taking into account the light collection and photocathode quantum efficiency, we find 1100  photons per $\alpha$ particle or  207 $\pm$ 30  photons/MeV. This is compatible with values of 220 photons per Mev in the range of  360-600 nm  measured at lower pressures \cite{Pan95}. 
     \\

    The time difference between  primary light and  avalanche light signals of alpha particles emitted from the cathode
provides an other measurement of the drift velocity.

 \subsubsection*{IV.2.2  Suppression of charge collection } 
\begin{figure}[htb]
 \vspace{1.cm}	
	\epsfxsize=8cm
	\epsfysize=8cm	
\centerline{\epsffile{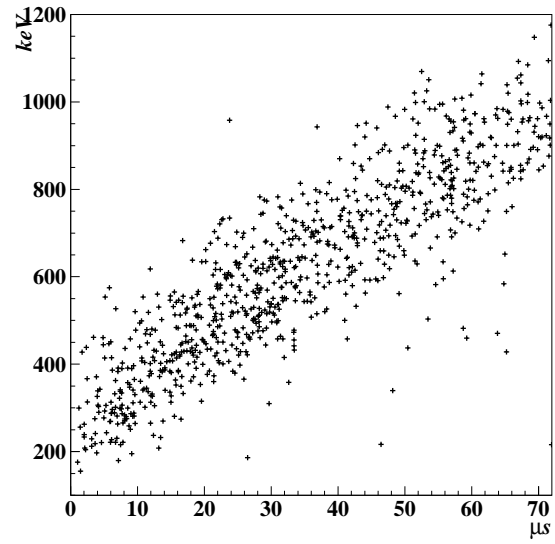}}
	\caption[]{Electron equivalent energy .vs. drift time (for $\alpha$ particles of 5.5 and 6 MeV).}
        \label{fi:alphag}        
\end{figure}

   The total charge (or avalanche light) collected for $\alpha$ particles is much smaller than for electrons of the same energy. 
   A suppression factor as high as 28 has been measured for $\alpha$ particle emitted close to the anode plane.  This factor however decreases when the drift time (i.e. distance from the anode) increases (Figure \ref{fi:alphag}). For alpha particles emitted from the cathode, the suppression factor is only 5.6. The charge suppression is very likely due to an screening effect in the avalanche region while the change of the suppression factor may be a consequence of electron diffusions along transverse(T) as well as longitudinal(L) directions.
   In the MUNU TPC, for a drift length of 1.62 m from the anode to the cathode, the dispersion $\sigma_{L}$ or $\sigma_{T}$ is 2 mm, about half of the range of $\alpha$ particles at 5 MeV. The smearing of the charge density along the drift path could than attenuate the charge space screening effect in the avalanche region. 
 \subsubsection*{IV.2.3  Radon identification and detection} 
\begin{figure}[htb]
 	\epsfysize=8cm	
\centerline{\epsffile{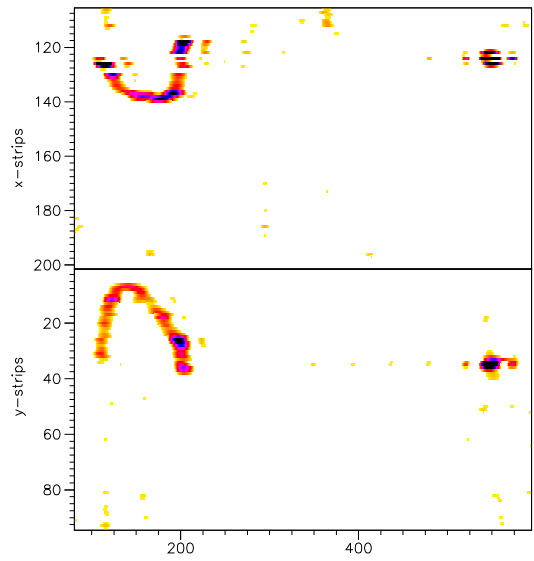}}
	\caption[]{A delayed $\beta$-$\alpha$ coincidence event}
        \label{fi:BiPo}
\end{figure}

\begin{figure}[ht]
        \vspace{1.cm}
	\epsfxsize=8cm
	\epsfysize=8cm
	\centerline{\epsffile{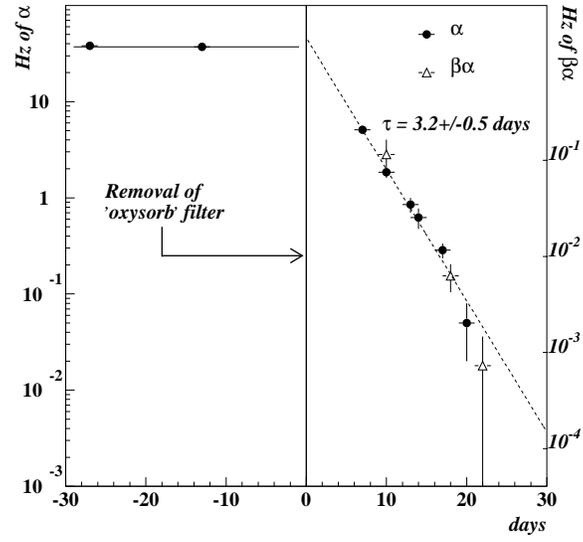}}
        \caption[]{Radon decay rates after removal of Oxysorb filter(see text).}
        \label{fi:decay} 
\end{figure}

Radon was introduced accidentally into the TPC by one of the Oxysorb \cite{oxys} purifiers used to clean the CF$_{4}$ from oxygen and water. A maximun rate of 37 s$^-1$ was measured.
	Alpha particles detected in the TPC came from the decay of $^{222}$Rn  (E$_{\alpha}$=5.48 MeV) and its radioactive daughters. In the radon $^{222}$Rn decay chain, the beta from $^{214}$Bi is followed by the $\alpha$ from $^{214}$Po(half life 160 $\mu$s). Such delayed coincidence events can be observed easily within the 80 $\mu$ s range of our flash ADC's. A correlated $\beta-\alpha$ event attributed to the $^{214}$Bi- $^{214}$Po decay cascade is shown in Figure \ref{fi:BiPo}.
When the Oxysorb was removed from the gas circuit, both the rates of $\alpha$ events and of correlated $\beta$-$\alpha$ events from $^{214}$Bi-$^{214}$Po  decay cascade  decreases with a period of 3.2 days compatible with that of  $^{222}$Rn decay period (Figure \ref{fi:decay}).
After the CF$_{4}$ gas change, residual $\alpha$ events were still observed at a steady level of $\sim$ 1 mBq/m$^3$ of $^{222}$Rn  in the gas and an activity of 17 $\mu$Bq/cm$^2$ from the cathode. For these last events, a well defined drift time and a well defined energy peak at 940 keV electron equivalent were observed.  These events can be attributed to $\alpha$ particles from $^{210}$Po decay, a radioactive daughter of $^{210}$Pb (half life 22 years). Radioactive daughters of $^{222}$Rn were very likely drifted to the surface of the copper cathode but apparently not deeply implanted into the metal. The cathode copper foil has now been replaced by a new one, the surface of which was chemicaly etched with radiopure chemicals before installment. The rate of $\alpha$ from the cathode  was reduced by a factor of 10.   
 
\subsection*{IV.3 Electrons}
  \subsubsection*{IV.3.1 Compton electrons}
\begin{figure}[ht]
\vspace{1.cm}	
	\epsfxsize=16cm
	\epsfysize=16cm	
\centerline{\epsffile{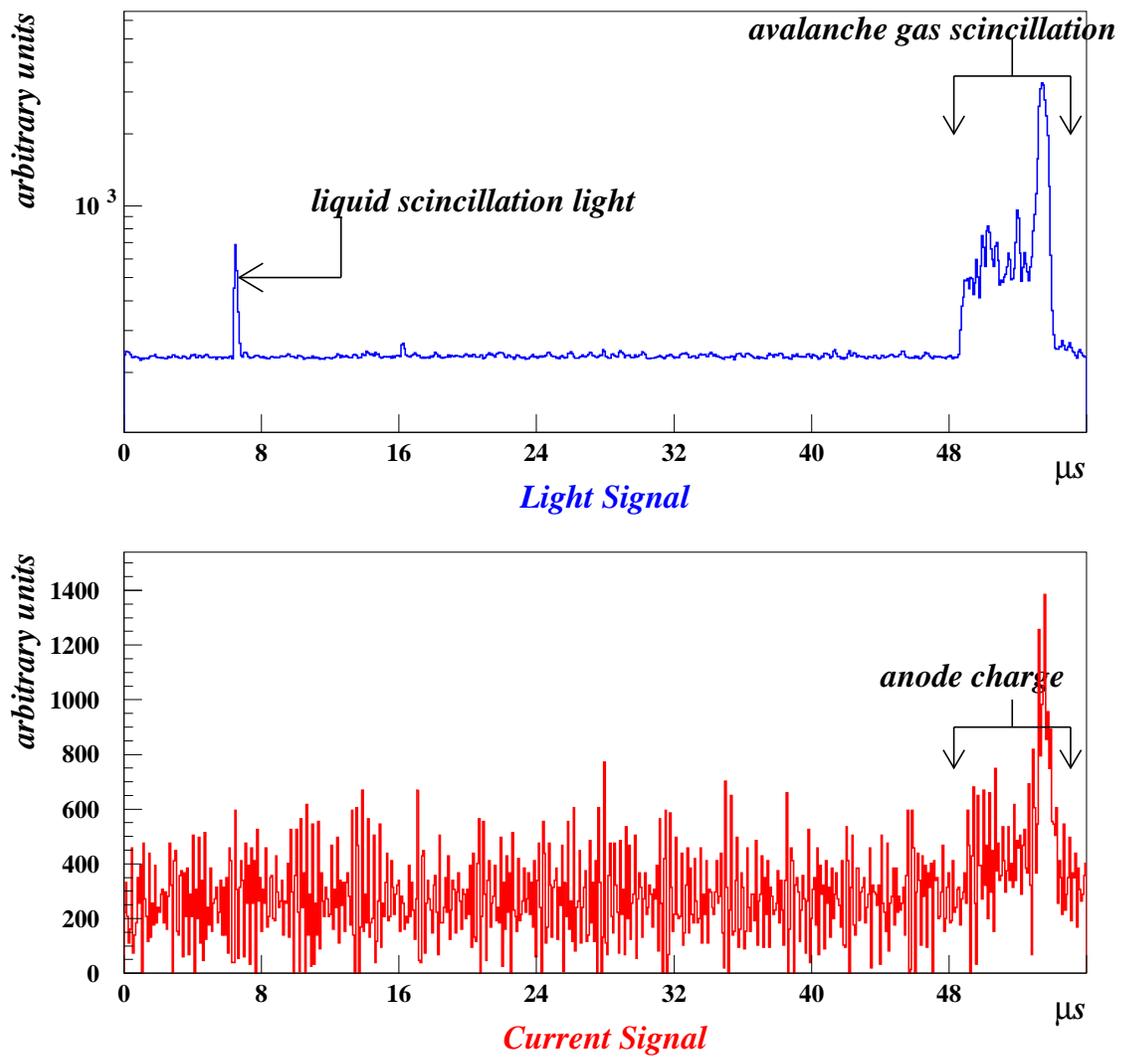}}
        \caption[]{ Waveforms of a Compton electron event}
        \label{fi:gammae}
\end{figure}

	Compton electron events generated within the TPC have in most cases an associated scattered $\gamma$ ray detected in the light collection vessel. The sequence for a typical event, shown in Figure \ref{fi:gammae} is the following: \\
- first, scintillation light due to the absorption of  the scattered $\gamma$ ray in the liquid scintillator is detected, \\
- then, when ionisation charges of the Compton electron have drifted to the anode plane, avalanche light is measured with a time delay proportional to the distance from the anode,\\	  
- almost simultaneously  electric current signals are measured on the anode and x-y strips.\\     
	The display of the event is completed by the reconstruction of the projections in the x-z/y-z and x-y planes. The localization of the gamma along the axis of the anti-Compton can be evaluated from the assymetry of the light collection(see section III.1).	
 The distribution of time differences of Compton electrons and associated scattered $\gamma$'s is flat along the TPC which indicates that the background is homogeneous along the TPC. A sharp drop corresponds to the end of the chamber. The drift velocity extracted this way is compatible with the muon result.
\subsubsection*{IV.3.2 Electrons from the anode plane} 
\begin{figure}[ht]
 \vspace{1.cm}	
	\epsfxsize=8cm
	\epsfysize=8cm	
\centerline{\epsffile{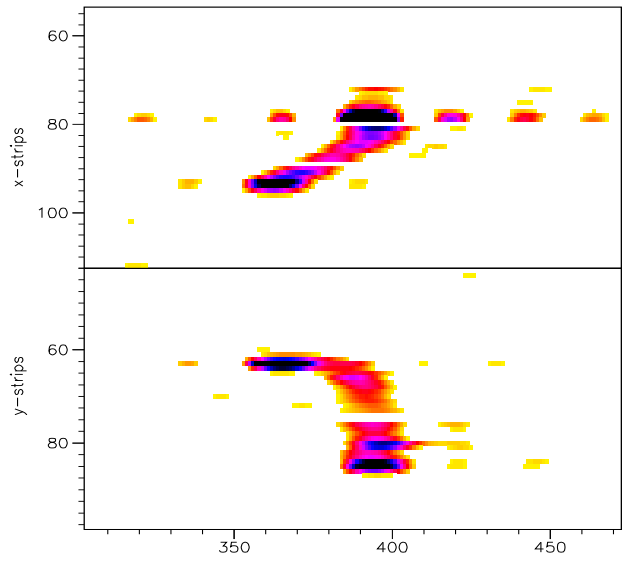}}
\vspace{1.cm}	
	\epsfxsize=8cm
	\epsfysize=8cm	
\centerline{\epsffile{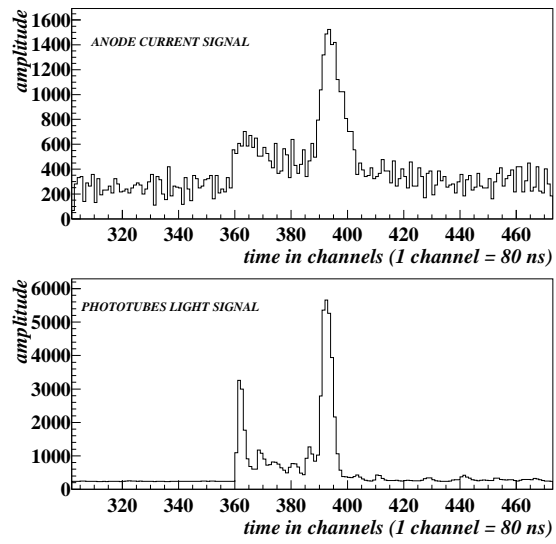}}
        \caption[]{A double blob event. The current and light signals
are shown together with the xz and yz projections.}
        \label{fi:twoblob}
\end{figure}

\begin{figure}[htb]	
	\epsfxsize=8cm
	\epsfysize=8cm		
\centerline{\epsffile{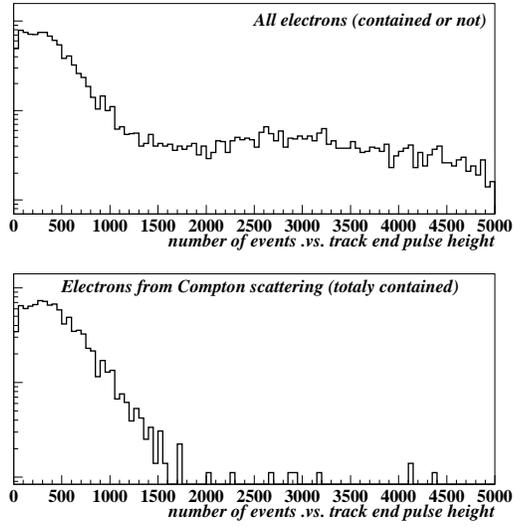}}
        \caption[]{Track end pulse height distributions (see text).}
        \label{fi:scrise}
\end{figure}

\begin{figure}[ht]
	\epsfxsize=8cm
	\epsfysize=8cm	
\centerline{\epsffile{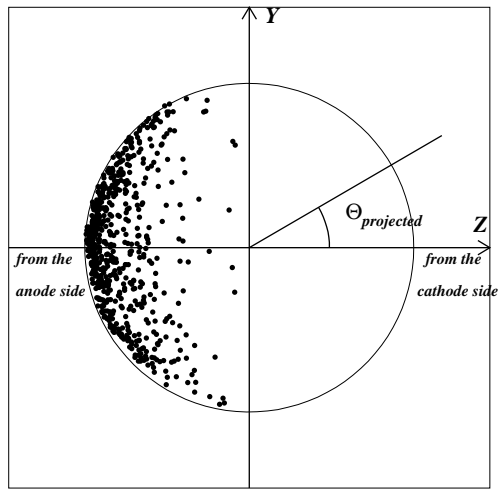}}
        \caption[]{Direction distribution of double blob events.}
        \label{fi:DA2blob}
\end{figure}

 	For such events, charges are created next to the anode wires. Thanks to a larger electric field and thus faster drift time between the grid and the anode the charges created in this region are collected much faster. The result is a charge blob at the end of the track on the anode side. If the electron stops inside the gas volume, the normal blob due to increased ionization before stopping will be seen in addition at the true end, resulting in a "double blob" event. Normal blobs have a higher integrated charge than anode blobs while anode blobs have charges accumulated in the first or the second time bins at the start of the track.  The avalanche light pulses are used to detect the sharp rise of pulse height and identify tracks originating from the anode plane.
A typical "double blob" event is shown in Figure \ref{fi:twoblob}.\\
A distribution of pulse heights for an unselected set of electron events is shown in Figure \ref{fi:scrise}  together with a distribution of Compton electron events which do not cross the anode. This last criteria is fullfilled by requiring an associated scattered $\gamma$ ray signal well separated from the avalanche signal. For a well ajusted cut (1500 mV or $\sim$ 20 keV), only 1.5$\%$ of normal Compton electron will be wrongly identified as originating from the anode. The direction distribution of  "double blob" events, projected in the yz plane(Figure \ref{fi:DA2blob}) shows that they are indeed dominantly originated from the anode side.  
\subsubsection*{IV.3.3 Electrons from lateral walls}
  Electrons emitted from lateral walls can be identified from their xy projections. In practice, a fiducial radius is defined to reject events having tracks touching the outermost x or y strips.
 Only electrons coming from the cathode wall are not distinguishable from contained electrons of the same direction.\\

\subsection*{IV.4 Contained single electron events }
\begin{figure}[ht]
	\epsfxsize=8cm
	\epsfysize=8cm	
\centerline{\epsffile{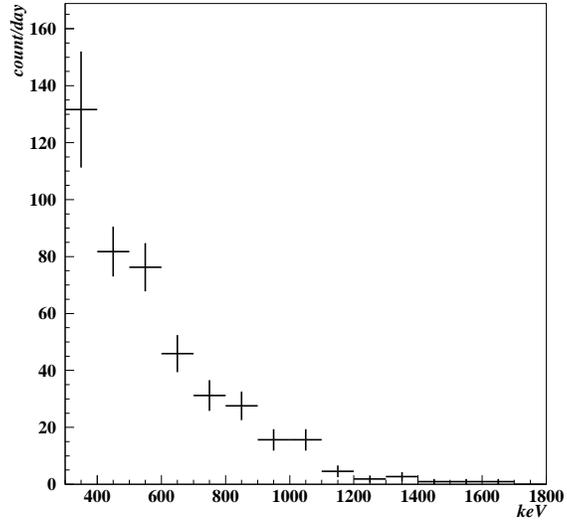 }}
        \caption[]{Energy spectrum of contained electron events.}
        \label{fi:ESTPC}	
\end{figure}

	The MUNU set-up is optimized for the detection of low energy single electron events in particular those from neutrino electron scattering. The intense electron anti-neutrino flux from the 2800MW power reactor at 18.6 m amounts to 1.2 10$^{13}$ $\nu$ /cm$^{2}$.s. The direction of neutrinos to the detector is well defined.($\pm$ 5$^o$)
 A neutrino candidate is a contained single recoil electron in the forward solid angle.  
 In data taking, we take dvantage of the properties of the light signals to 
apply hardware and online cuts to reject undesired types of events: 
muon associated events, "micro electric flashes", alpha particles and Compton electrons. Electron tracks down to 300 keV have been measured.
The gross counting rate in the TPC (70 s$^{-1}$) is dominated by cosmic muons, Compton electron events amounting to 0.14 s$^{-1}$. 
  After successive cuts: muon rejection, fiducial volume cuts (events touching the lateral walls or the anode plane), then anti-Compton rejection( 100 keV threshold), the counting rate is 780 day$^{-1}$. The counting rates are shown in Table \ref{tb:rates}. The recoil energy spectrum of contained electron events is shown in Figure \ref{fi:ESTPC}. \\  
  For neutrino electron scattering events induced by far away reactor,  a  solid angle cut can further be applied. For a forward angle confined within ($\pi$/4), the residual counting rate is 97 day$^{-1}$ per 11kg  of CF$_{4}$ at an energy threshold of 300 keV or  8.8 (kg.MeV.day)$^{-1}$.\\
 The recoil energy spectrum of contained electron events does not show any precise structure.  Very likely, the remaining background does not have a dominant origine but may have different components: $\gamma$ contaminants of the walls and the electrodes of the TPC or single electron from the gas itself (cosmic activation or presence of radioactive gas). 
 This is under further investigations. 
\section*{\bf V. Angular and energy resolution}   
 \subsection*{V.1 Energy resolution} 
 The energy resolution of a wire chamber in proportional regime
usually depends on the primary ionization statistics and mainly on
the single electron avalanche response spectrum, which  
dominates in the present case of pure CF$_4$ through a Polya parameter 
$\theta=0.21$ already measured in ref \cite{Va'vra}.\\ 
Moreover this gas reveals a strong attachment for values of 
the reduced field between 40 and 140 V.cm$^2$. While no attachment occurs
during the drift phase due to the low value of drift field, nearly $98\%$ of
the drifted electrons are attached in the early stage of the avalanche,
affecting the resolution \cite {Anderson}.\\ 
For a primary deposition of 300 keV in the TPC only 2$\%$ of the primary 
ionization electrons will survive and reach the multiplication phase 
of the avalanche, the same number as for a 6 keV deposition and
no attachment; the fluctuations should then be of the same order of magnitude.\\  
However the energy of a track is not measured through one single avalanche,
but through successive ones (up to 100 independent avalanches for a 300 keV 
track, depending on track length, energy loss per unit length and angle 
in respect to the drift direction). The overall fluctuations are then reduced 
when compared to a single measurement.
A complete calculation including the previous parameters and simulated tracks
with energy loss tables predicts a resolution of $\sigma=10.1\%$  at 640 keV 
in agreement with measurements.\\
Due to the different contributions, the resolution is not expected to follow 
a simple $\sqrt E$ law; a fit to the simulations at different energies rather 
shows an empirical  $E^{0.7}$ law, in better agreement with the low energy part 
of the Compton spectra.\\

\subsection*{V.2 Angular resolution and incident energy reconstruction} 

The angular resolution has been measured through $\gamma-e^-$ Compton 
scattering from various sources. The diffused $\gamma$ is measured in the 
scintillator, the recoiling electron track in the TPC.
The time difference between the $\gamma$ and the drifted electron gives 
the absolute localization along the drift direction, allowing 
(together with the XY track projection) a full 3D determination 
of the vertex of the interaction and then the exact incident $\gamma$ ray 
direction.\\ 
The electron scattering angle is obtained by fitting the direction of 
the first centimeters of the track since the angular resolution is mainly
limited by multiple scattering in the first few centimeters of sampled track.
In the present case the electron track recognition, the vertex determination 
and the direction were obtained via a fully automatic image processing which 
will be described elsewhere.\\

\begin{figure}[ ]
	\epsfxsize=8cm
	\epsfysize=8cm	
	\centerline{\epsffile{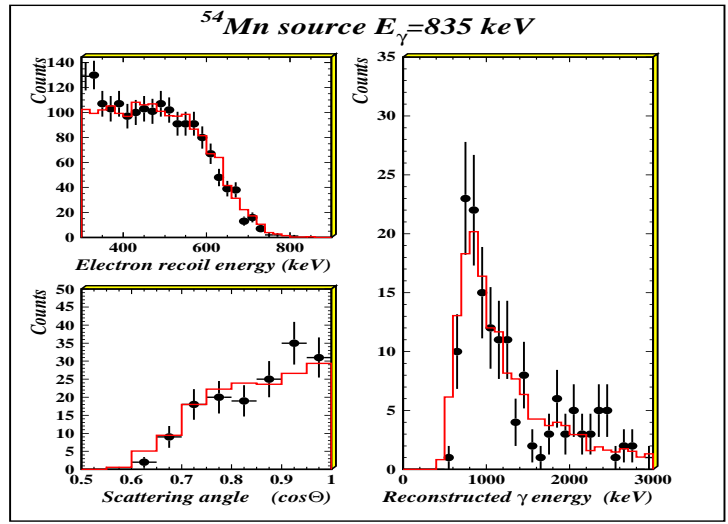 }}   
      \caption[]{Initial gamma energy reconstruction(see text).}
      \label{fi:Resolang}
\end{figure}

Figure \ref{fi:Resolang} shows the reconstructed 835 keV $\gamma$ ray from a $^{54}$Mn source 
measured at 3 bars with a recoiling electron energy threshold of 300 keV. The $\sigma$ of the reconstructed energy peak at 835 keV is 220 keV, although the shape is far from a gaussian shape. 
The overall angular resolution averaged over the Compton recoil spectrum above 300 keV threshold is $\sigma_{\theta}$ = 39.7$^o\pm$ 2.2. 
  To our knowledge this is the first time that a $\sim$ 1 MeV photopeak is reconstructed by measuring the Compton scattering in a gaseous detector.

This result is encouraging in view of using a gas TPC to detect solar neutrinos. It demonstrates that the $^7$Be solar neutrino peak at 862 keV could be reconstructed even at 3 bars by measuring the neutrino-electron scattering with a detector based on ths same principe as ours. By lowering the pressure to 1 bar, we shoud be able to measure the pp solar neutrino which has the end-point at 420 keV. 
   
\section*{\bf  VI. Conclusions} 
 
The MUNU detector is a 1 m$^{3}$ TPC enclosed within a 10 m$^{3}$ light collection vessel. It has been built to study of $\overline{\nu}_{e}e^-$ scattering at the Bugey reactor and to set a lower experimental limit on the neutrino magnetic moment. It is particulary dedicated to detect low energy electrons and to reconstruct their energy and direction. 
   
One of the originalities of MUNU is the presence of the external light 
collection vessel. It sees the light produced by ionizing particles in 
its scintillator. This allows to eliminate cosmics related events as 
well as Compton events occuring in the TPC. In addition the primary 
scintillation light produced by alphas is well resolved, which allows 
to identify these particles. For all these events the measurement of 
the primary light provides an absolute determination of the track 
position along the TPC axis. 

The primary light from minimum ionizing particles is too weak to be 
seen, but the light emitted at the anode during the amplification 
process gives a strong signal. It complements that from the anode 
signal, having even lower noise. It is very useful in characterising 
the topology of tracks, which will allow for a powerful event 
selection when analysing the data in terms of neutrino scattering. 
 Installed within low activity shieldings, the detector is now collecting neutrino induced electron events at the Bugey reactor at a recoil electron threshold of 300 keV.\\
    The result of the reconstruction of the incident energy obtained with the 835 keV $^{54}$Mn source is encouraging and it shows that the spectroscopy of low energy neutrino is attainable. The background has to be however further suppressed.
    The ultimate goal is to use an upgraded MUNU detector underground, where the background conditions are optimal, and to look for solar neutrinos. \\
    
Aknowledgement\\
This work has been supported by the Institut National de Physique Nucl\'eaire et de Physique des Particules (IN2P3/CNRS), INFN, le Fond National Suisse. \\ 
 We are grateful to the staff of the Bugey nuclear power plant (EDF) for their hospitality and help.\\ 
 The support of the technical staff of the participating laboratories, in particular that of C. Barnoux, B. Guerre-Chaley, L. Eraud, M. Marton, R. Blanc, D. Schenker, J.M. Vuilleumier, D. Filippi, D. Maniero, M. Negrello  is greatly aknowledged\\
  G.Bagieu, R. Brissot, J.M. Laborie,  G. Gervasio took part in the early phase of the experiment, we thank for their contributions. 

\end{multicols}
\clearpage
\begin{table}[]
\caption{Counting rates in the MUNU detector }
\label{tb:rates}
\begin{tabular}{||c|c||}
cuts&rates\\
\hline 
T $>$ 150 keV & 70/s\\ 
T $> $300 keV and no muon veto  & 0.35/s\\ 
within fiducial volume & 0.15/s\\ 
no anti-Compton veto  & 778/day\\
forward angle within $\pi$/4 & 97/day\\ 
\end{tabular}
\end{table}
\newpage 
\end{document}